\documentclass[aps,twocolumn,showpacs,preprintnumbers,amsmath,amssymb,floatfix]{revtex4}
\usepackage{graphicx}
\usepackage{latexsym}
\usepackage{verbatim,times,bbm}
\usepackage{color}

\begin{document}

\newcommand{\REV}[1]{\textbf{\color{red}#1}}

\title{Robustness against parametric noise of non ideal holonomic gates}
\author{Cosmo Lupo}
\author{Paolo Aniello}
\author{Mario Napolitano}
\affiliation{Universit\`a di Napoli ``Federico II'' and INFN,
Sezione di Napoli, via Cintia I-80126 Napoli, Italy}
\author{Giuseppe Florio}
\affiliation{Universit\`a di Bari and INFN, Sezione di Bari, I-70125
Bari, Italy}

\begin{abstract}

Holonomic gates for quantum computation are commonly considered to
be robust against certain kinds of parametric noise, the very
motivation of this robustness being the geometric character of the
transformation achieved in the adiabatic limit. On the other hand,
the effects of decoherence are expected to become more and more
relevant when the adiabatic limit is approached. Starting from the
system described by Florio {\it et al.}\ [Phys.~Rev.~A {\bf 73},
022327 (2006)], here we discuss the behavior of non ideal holonomic
gates at finite operational time, i.e., far before the adiabatic
limit is reached. We have considered several models of parametric
noise and studied the robustness of finite time gates. The obtained
results suggest that the finite time gates present some effects of
cancellation of the perturbations introduced by the noise which
mimic the geometrical cancellation effect of standard holonomic
gates. Nevertheless, a careful analysis of the results leads to the
conclusion that these effects are related to a \emph{dynamical}
instead of geometrical feature.

\end{abstract}


\maketitle

\section{\label{intro}Introduction}

One of the most important challenges for the realization of quantum
information tasks is the implementation of quantum logic gates that
are \emph{robust} against unwanted
perturbations~\cite{Nielsen_Chuang,Casati}. Two kinds of
perturbations with qualitatively different features can be
distinguished: the first kind has a purely \emph{quantum} nature,
and it is induced by the interaction of the quantum system
implementing the logic gate with the environment; the second kind
has instead a \emph{classical} nature, and it is caused by the
presence of instrumental noise in the `external parameters' used to
control the system. The unwanted interaction with the environment is
the source of the phenomenon known as quantum
\emph{decoherence}~\cite{Breuer}. The effects of this interaction
can be modeled by means of suitable `master equations' (i.e.\
evolution equations) for the density matrix of the quantum system
implementing the logic gate; at least in the Markovian regime
\footnote{It is worth stressing  that in the limit for the
operational time tending to zero the Markovian approximation cannot
be applied, in principle --- see, for instance, ~\cite{nonMarkov}
--- and one cannot claim, in general, that environmental noise can
be avoided by simply decreasing the operational time.}, they are
negligibly small if the \emph{operational time} of the logic gate is
short enough. The classical perturbations stem from an unavoidable
noisy component intrinsic in the external driving fields (e.g.\
laser beams~\cite{NIST}) that can be usually regarded as classical
fields; hence, it is essentially due to instrumental instability.
The effects of these perturbations can be evaluated by studying
standard (non-autonomous) Schr\"odinger equations where the
instrumental noise is taken into account by suitably modeling the
noisy components of the classical parameters (e.g.\ the field
amplitude) associated with the external driving fields.

Among the several strategies for realizing quantum logic gates
discussed in the literature, a prominent position is held by
\emph{holonomic gates}. They were first proposed by Zanardi and
Rasetti~\cite{Zanardi_Rasetti} (see also Ref.~\cite{PachosPRA61}),
and rely on the theory of holonomy and of the associated holonomy
groups in principal fiber bundles~\cite{Nakahara}, a subject which
is familiar to theoretical physicists due to the central role played
in gauge theories~\cite{Marathe} and in the well-known phenomenon of
abelian~\cite{Berry} and non-abelian~\cite{WZ} adiabatic phases.
Actually, a holonomic gate can be regarded as a straightforward
application of the theory of non-abelian adiabatic phases to quantum
computation.

Since the very beginning, holonomic gates were considered to be
intrinsically robust against classical noise~\cite{Pachos}, thanks
to the geometric features of holonomy in Hilbert bundles. As we will
briefly recall below, three main ingredients are needed in order to
realize such holonomic gates.\\
The first ingredient is a suitable physical system described by a
quantum Hamiltonian depending on some set of parameters, these
parameters being associated with the external (classical) driving
fields that are assumed to be experimentally controllable functions
of time; the unavoidable instrumental instability (stochastic noise)
affecting the driving fields is the source of the classical noise
--- we will call it \emph{parametric noise}, in the following
--- that has been mentioned above.\\
The second ingredient consists in selecting a suitable eigenspace
of the given Hamiltonian --- an eigenspace depending smoothly on
the external parameters, hence actually an iso-degenerate family
of eigenspaces; let us call them the family of \emph{relevant
eigenspaces} --- and in fixing in the parameter space an `initial
point' and a loop through this point. To such a loop corresponds
an excursion of the parameter-dependent Hamiltonian (hence, of its
eigenprojectors) and a certain \emph{ideal unitary transformation}
in the \emph{encoding eigenspace}, namely, that particular
relevant eigenspace fixed by the initial (and final) point of the
loop in the parameter space. This ideal transformation is
determined by \emph{Kato's adiabatic evolutor} associated with the
given Hamiltonian and with the chosen loop in the parameter space,
and it has a simple geometric interpretation as a holonomy
phenomenon (geometric phase). The ideal unitary transformation
plays a central role in Kato's formulation of the adiabatic
theorem~\cite{Kato} applied to our context. Indeed, the external
parameters are controllable functions of time and in the
\emph{adiabatic limit} --- i.e., in the limit where the loop in
the parameter space is covered in a operational time tending to
infinity
--- the \emph{real evolution over the operational time},
determined by the given physical Hamiltonian, becomes \emph{cyclic}
in the encoding eigenspace and, apart from an irrelevant overall
`dynamical phase factor', \emph{coalesces in this subspace with the
ideal unitary transformation}. We stress that the ideal unitary
transformation should be thought, in our context, as an \emph{ideal
quantum gate} whose behavior can be, in general, only approached by
a \emph{non-ideal quantum gate} corresponding to the real
evolution over a suitably large, but finite, operational time.\\
Accordingly, the third ingredient is the choice of a suitable
operational time --- which will be called \emph{balanced working
time}, in the following --- for the real quantum gate. This time
span must be short enough to achieve a fast quantum computer and
to avoid the ravages of decoherence, but long enough to justify
the adiabatic approximation (i.e.\ to approach the behavior of the
ideal quantum gate) which is at the root of the appearing of
geometric phases \footnote{We recall that geometric phases arise
also in the context of (non-adiabatic) cyclic
evolutions~\cite{Aharonov,Anandan}, but only \emph{adiabatic}
phases are relevant for our purposes.}. Hence, a balanced working
time is determined by a touchy trade-off between two competing and
not necessarily compatible demands.

The problem of robustness of holonomic gates against parametric
noise has been studied in both the abelian and non-abelian case with
different
approaches~\cite{DeChiara_Palma,Solinas_Zanardi,Zu_Zanardi}. In
these papers, the effects of random perturbations of the control
parameters are considered. It is worth noticing, however, that such
effects are evaluated with the adiabatic limit already being
performed, thus essentially confirming quantitatively the standard
qualitative \emph{geometric argument} usually adopted to support the
robustness of holonomic gates, argument which will be recalled later
on.

As holonomic gates are generally considered to be \emph{a priori}
robust against parametric noise, attention has mainly focused on
the study of decoherence
effects~\cite{Carollo,Carollo2,Fuentes,Parodi} and on the
possibility of partially suppressing them~\cite{Wu_Zanardi_Lidar}.
These investigations show that, for certain physical systems and
for certain models and regimes of the coupling with the
environment, one is able to estimate the typical time-scale within
which the effects of decoherence can be neglected. Hence one can
determine, in principle, a balanced working time for these
systems. At this point one should actually \emph{check} whether
this balanced working time guarantees a suitable robustness of the
quantum gate against parametric noise, namely, whether the effects
of this kind of noise on the fidelity of the non-ideal quantum
gate with respect to the ideal one can be neglected or not.

Recently, a new ingredient has been proposed for the
implementation of a holonomic quantum gate~\cite{Florio_1} (see
also~\cite{Florio_2,Florio_3}) where the authors have observed
--- for the model of a ion-trap quantum gate proposed by Duan {\it
et al.}~\cite{Cirac} --- the existence of a \emph{optimal working
time}, namely, of a specific operational time for which the
non-ideal (i.e.\ finite-time) gate behaves \emph{exactly} as the
ideal (i.e.\ adiabatic) gate; they show, furthermore, that over
the optimal working time the effects of the environment are
negligible. Thus, such a optimal working time turns out to be also a balanced working time.\\
We stress that, anyway, the fact that the non-ideal gate behaves, in
correspondence to the optimal working time, as the ideal one cannot
be used to rule out the influence of parametric noise on the base of
the standard geometric argument. Indeed, one should not expect that,
perturbing the loop in the parameter space, the non-ideal gate will
still mimic the behavior of the ideal one. Hence one cannot apply,
in principle, the standard geometric argument to support the
robustness of this kind of holonomic gate against parametric noise.
A critical analysis of this simple, but somehow subtle, issue is the
main aim of the present contribution.

In conclusion, we think that the impact of parametric noise on
holonomic gates is still an open problem and one is not legitimated,
in general, to state the robustness of non-ideal holonomic gates
against this kind of perturbations on the base a generic geometric
argument. In our present contribution, we will try to illustrate
this assertion by means of quantitative arguments, focusing on the
ion-trap model proposed by Duan {\it et al.}~\cite{Cirac}. Even if
other models have been proposed in the literature~\cite{Falci}, the
model of Duan {\it et al.}\ is probably the one most extensively
studied also with reference to different physical systems, as
Josephson junctions~\cite{Faoro} and semiconductor quantum
dots~\cite{Solinas}, and can be regarded as a reference point for
the subject.

The paper is organized as follows. In Sec.~\ref{case.study} the
model Hamiltonian is introduced which will serve as a case study. In
Sec.~\ref{noise.model} the behavior of the considered system in
presence of several models of parametric noise is discussed. In Sec.
\ref{kritik} the obtained results are analyzed and commented.
Conclusions and remarks are presented in Sec.~\ref{conclude}.

\section{\label{case.study}A case study}

\begin{figure}
\centering
\includegraphics[width=0.48\textwidth]{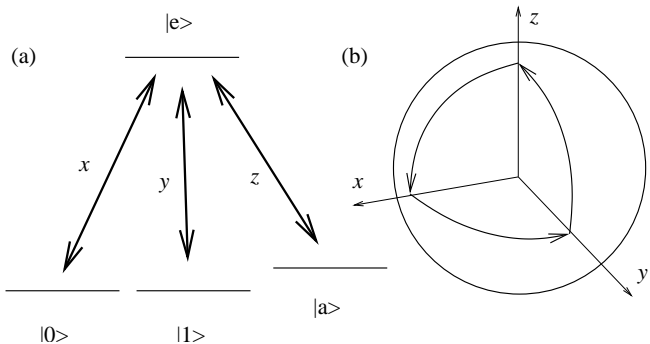}
\caption{Structure of the atomic levels and resonant lasers (a);
unperturbed loop (\ref{path}) in the parameter manifold (b).}
\label{loop}
\end{figure}

As a case study, here we consider the single-qubit non Abelian gate
that was proposed in~\cite{Cirac}. The model under consideration can
be physically realized as, for instance, a trapped ion with two
degenerate ground (or metastable) states $|0\rangle$ and $|1\rangle$
which play the role of the computational basis. A quasi degenerate
ancillary state $|a\rangle$ and an excited state $|e\rangle$ are
also needed (the scheme is drawn in Fig.~\ref{loop}(a)). The low
energy states are supposed to be independently coupled with the
excited state, such that the interaction picture Hamiltonian in the
rotating frame reads as follows:
\begin{equation}\label{Ham}
H(\mathbf{r}) = H(x,y,z) =
|e\rangle \left[ x \langle 0| + y \langle 1 | + z \langle a | \right] + \mathrm{h.c.}\;.
\end{equation}
The, in general complex, parameters $x, y, z$ are related to three
independent Rabi frequencies corresponding to, in general de-tuned,
laser beams with different energies and polarization. In an ideal
experiment, however, the laser beams are assumed to be resonant with
the corresponding transitions and the parameters are constrained to
take values on a two-sphere. It is thus convenient to introduce
polar coordinates:
\begin{eqnarray}\label{r_path}
\left\{
\begin{array}{ccc}
x & = & \Omega \sin{\vartheta} \cos{\varphi} \\
y & = & \Omega \sin{\vartheta} \sin{\varphi} \\
z & = & \Omega \cos{\vartheta}
\end{array}\right.\;.
\end{eqnarray}
The spectrum of (\ref{Ham}) is threefold: $\sigma = \{ 0 ,\pm \Omega
\}$, with the null eigenvalue which is doubly degenerate. The two
degenerate eigenstates with vanishing energy can be chosen as
follows:
\begin{equation}\label{basis}
\begin{split}
|\psi_0\rangle & = \cos{\vartheta} \left( \cos{\varphi} |0\rangle +
\sin{\varphi} |1\rangle \right) - \sin{\vartheta} |a\rangle, \\
|\psi_1\rangle & = -\sin{\varphi} |0\rangle + \cos{\varphi}
|1\rangle.
\end{split}
\end{equation}

An analysis of the holonomy associated to the Hamiltonian
(\ref{Ham}) in correspondence with the doubly degenerate subspace
shows that a closed path with starting point $\vartheta = 0$
corresponds to a non Abelian holonomy $W = \exp{\left[- i \sigma_y
\omega \right]}$, where $\sigma_y = -i \left( |0\rangle\langle 1| -
|1\rangle\langle 0| \right)$ is the Pauli matrix in the
computational space and $\omega$ is the solid angle spanned by the
parameter $\vartheta(s)$ and $\varphi(s)$. This geometric character
of the dynamics in the adiabatic limit is at the heart of the usual
argument in favor of the robustness of holonomic quantum
computation. A stochastic noise in the control parameter can modify
the details of the loop but, for a sufficiently great number of
cycles of the noise during the system evolution, the fluctuations in
the swept solid angle are consider to become negligible
(see~\cite{DeChiara_Palma,Solinas_Zanardi} and references therein).

Here we consider the closed path in the parameter manifold that was
studied in~\cite{Florio_1}. For $s \in [0,1]$ we take (see
Fig.~\ref{loop}(b)):
\begin{equation}\label{path}
\begin{split}
\vartheta(s) = & \left\{\begin{array}{ll}
3 s \pi/2 & s \in [0,1/3] \\
\pi/2 & s \in [1/3,2/3] \\
3 \pi/2 \left( 1 - s \right) & s \in [2/3,1]
\end{array}\right.
\\
\varphi(s) = & \left\{\begin{array}{ll}
0 & s \in [0,1/3] \\
3 \pi/2\left(s-\frac{1}{3}\right) & s \in [1/3,2/3] \\
\pi/2 & s \in [2/3,1]
\end{array}\right.
\end{split}
\end{equation}
The solid angle related to the loop (\ref{path}) is $\omega =
\pi/2$, hence the corresponding holonomic gate is $W = - i
\sigma_y$. As was observed in~\cite{Florio_1}, the remarkable
property of this path is that it presents perfect revivals of the
gate fidelity at finite operational time. The same behavior was
predicted for all the loops constructed by moving from the north
pole to the equator through a meridian and back to the north pole
through another meridian with piecewise constant velocity. In the
case of the loop (\ref{path}) there is a perfect revival of fidelity
in correspondence of the operational times:
\begin{equation}
\tau^*_k = \frac{3\pi}{2\Omega} \sqrt{16k^2-1}, \ \
k=1,2,\dots\label{opttimes}
\end{equation}
In the following we are mostly concerned with the first optimal
operational time $\tau^*_1$.

To conclude this section we notice that a geometric phase appears in
correspondence to a non adiabatic {\it cyclic}
dynamics~\cite{Aharonov,Anandan}. In particular, for our case study,
it happens that, in correspondence to an optimal operational time,
the evolution becomes cyclic and the acquired geometric phase is
equal to the adiabatic holonomy.

\section{\label{noise.model}Models of noise}

In order to study the robustness of non ideal holonomic gates, we
consider the response of the system under parametric noise in the
ideal loop (\ref{path}). In order to quantify the robustness of the
gate, the noisy finite time evolution of the system is solved with
analytic or numerical methods and the average gate fidelity is
calculated. In the following, several models of noise are taken into
account: in Sec.~\ref{sin} we consider the response of the system
under a monochromatic perturbation of the three Rabi frequencies in
(\ref{Ham}); in Sec.~\ref{sphere} we consider a model of noise
expressed by a random step function in the angular variables
(\ref{r_path}) on the sphere; finally, in Sec.~\ref{random} we
discuss the response of the system under a random perturbation in
the three Rabi frequencies.

\subsection{\label{sin}Monochromatic perturbation}

In this section we consider the behavior of the system in presence
of a small perturbation in the parameters which can be viewed as a
\emph{probe} function used to test the stability of the gate, in
particular we concentrate our attention to the case of a
monochromatic probe function. A generic noisy path can be written as
follows:
\begin{equation}\label{noisy_path}
\mathbf{r}_{ \mathrm{n}}(t) = \mathbf{r}(t) +
\boldsymbol{\epsilon}(t), \ \ t \in [0,\tau],
\end{equation}
where the vector $\mathbf{r}(t)$ describes the unperturbed loop and
$\boldsymbol{\epsilon}(t)$ is a three component vector including the
perturbation of the path.
We have chosen a monochromatic perturbation at frequency $\eta$ and
considered a noisy path obtained from (\ref{r_path}) and
(\ref{path}):
\begin{eqnarray}\label{monoc_noise}
\left\{
\begin{array}{ccc}
x_\mathrm{n}(s;\eta,\epsilon_\eta,\tau,\phi_1) & = & x(s) + \epsilon_\eta e^{i\eta \tau s + i\phi_1} \\
y_\mathrm{n}(s;\eta,\epsilon_\eta,\tau,\phi_2) & = & y(s) + \epsilon_\eta e^{i\eta \tau s + i\phi_2} \\
z_\mathrm{n}(s;\eta,\epsilon_\eta,\tau,\phi_3) & = & z(s) +
\epsilon_\eta e^{i\eta \tau s + i\phi_3}
\end{array}
\right.\;,
\end{eqnarray}
where $\mathbf{r}_\mathrm{n}(s) \equiv
(x_\mathrm{n}(s),y_\mathrm{n}(s),z_\mathrm{n}(s))$,
${\boldsymbol{\phi}} \equiv (\phi_1, \phi_2, \phi_3)$ are random
phases uniformly distributed in $[0,2\pi)$ and $\epsilon_\eta$ is
the strength of the noise (chosen to be equal for the three
component). Notice that this model of noise acts on both the
amplitude and the de-tuning of the lasers. From (\ref{monoc_noise})
it is clear that at finite operational time the perturbation does
\emph{not} reduces to a geometric perturbation of the loop in the
parameters space since the perturbed path itself depends on the
operational time. In presence of noise, different values of the
operational time $\tau$ correspond to different loops in the
parameters manifold.

For given values of $\eta, \epsilon_\eta, \tau$ and
$\boldsymbol{\phi}$, we consider the solution of the Schr\"odinger
equation
\begin{equation}\label{noisy_SE}
V_\tau'(s;\eta,\epsilon_\eta,\boldsymbol{\phi}) = -i \tau
H(\mathbf{r}_\mathrm{n}(s))
V_\tau(s;\eta,\epsilon_\eta,\boldsymbol{\phi}), \ \ s \in [0,1].
\end{equation}
where, in presence of noise, the re-scaled Hamiltonian
$H(\mathbf{r}_\mathrm{n}(s))$ depends on $\tau$ too. Since we are
mainly interested in the transformation emerging at the end of the
loop, we set $V_\tau(\eta,\epsilon_\eta,\boldsymbol{\phi}) \equiv
V_\tau(1;\eta,\epsilon_\eta,\boldsymbol{\phi})$.

Notice that, for all practical purposes, taking the average on the
random phases corresponds to the action of the completely positive map
\begin{equation}\label{CPM}
\rho \longrightarrow \mathcal{E}(\rho) = \frac{1}{(2\pi)^3} \int
d{\boldsymbol{\phi}} V_\tau(\eta,\epsilon_\eta,\boldsymbol{\phi})
\rho V_\tau(\eta,\epsilon_\eta,\boldsymbol{\phi})^\dag\;.
\end{equation}
This completely positive map has to be compared with the ideal
adiabatic unitary dynamics, to do that, we have evaluated the
average gate fidelity
\begin{equation}
F = \int d\psi \langle\psi| W^\dag
\mathcal{E}(|\psi\rangle\langle\psi|) W |\psi\rangle\;,
\end{equation}
where $d\psi$ indicates the normalized Fubini-Studi metric on pure
states. This has been computed by means of the formula (see
Ref.~\cite{Nielsen}):
\begin{equation}\label{pauli_fid}
F = \frac{1}{3} + \frac{1}{12} \mathrm{tr}\left[ W W^\dag
\mathcal{E}(P_0(0)) \right] + \frac{1}{12} \sum_{j=1}^3
\mathrm{tr}\left[ W \sigma_j W^\dag \mathcal{E}(\sigma_j) \right]
\end{equation}
where $\sigma_j$ are the Pauli matrices in the computational subspace.

For several values of $\eta, \epsilon_\eta$ and $\boldsymbol{\phi}$,
Eq.~(\ref{noisy_SE}) is numerically solved using the relation:
\begin{equation}\label{algorithm}
V_\tau(\eta,\epsilon_\eta,\boldsymbol{\phi}) =
\lim_{N\rightarrow\infty} \stackrel{\leftarrow}{\prod}_{k = 0 \dots
N} \mathrm{exp} \left[ - i \tau H(\mathbf{r}_\mathrm{n}(k/N))
\frac{1}{N} \right]
\end{equation}
Where $\overleftarrow{\prod}$ stands for the path ordered product.
The effective completely positive map (\ref{CPM}) is evaluated
taking the average over $50$ random choices of the phases
$\boldsymbol{\phi}$.
\begin{figure}
\centering
\includegraphics[width=0.5\textwidth]{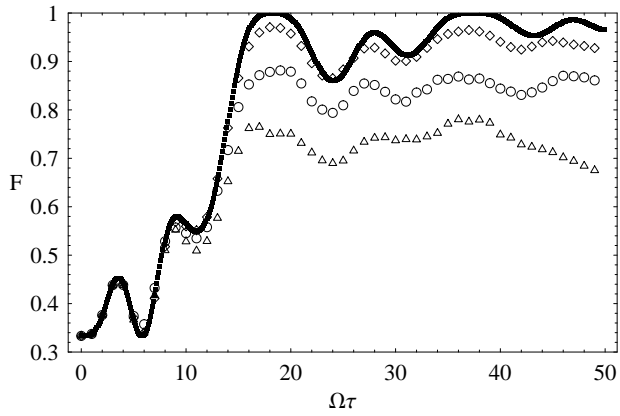}
\caption{Average gate fidelity as a function of the adimensional
operational time $\Omega\tau$ for several noise frequencies for the
model in Sec.~\ref{sin}. Black boxes: $\epsilon_{\eta} = 0$;
circles: $\epsilon_{\eta} = 0.1 \Omega, \eta = 0.1 \Omega$;
triangles: $\epsilon_{\eta} = 0.1 \Omega, \eta =0.2 \Omega$;
squares: $\epsilon_{\eta} = 0.1 \Omega, \eta = 0.3 \Omega$.}
\label{F(T)}
\end{figure}
Figure \ref{F(T)} shows the estimated gate fidelity
(\ref{pauli_fid}) plotted as a function of the adimensional
operational time $\Omega \tau$, for several values of the noise
amplitude and frequency. The unperturbed dynamics corresponds to
$\epsilon_{\eta} = 0$ and can be compared with the analytical
results in~\cite{Florio_1}, it exhibits perfect revivals of the
average gate fidelity at finite time, in particular the first
optimal operational time is $\Omega \tau^*_1 \simeq 18.25$. The
numerical results show that the pattern of gate fidelity as a
function of the operational time can be completely different in
presence of noise.

The average gate fidelity at the first optimal operational time
$\tau^*_1$ in presence of parametric noise is plotted in
Fig.~\ref{F(v,e)} as a function of both amplitude and frequency of
the noise. This plot suggests that the gate is indeed robust also
for rather large noise amplitude ($\epsilon_{\eta} = 0.4 \Omega$).
It is worth to notice that this is true unless the perturbation
frequency is in a particular range approximatively about $\eta
\simeq 0.15 \Omega$. The presence of a typical frequency scale in
the pattern of the fidelity is a feature that will be reencountered
in the other models of noise considered below.
\begin{figure}
\centering
\includegraphics[width=0.48\textwidth]{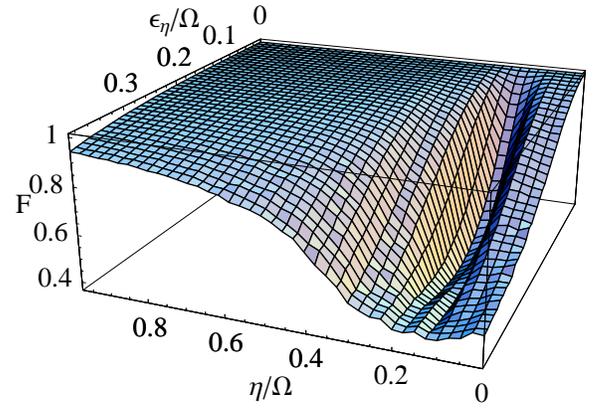}
\caption{(Color on line.) Average gate fidelity at the first optimal
operational time as a function of adimensional noise frequency
($\eta/\Omega$) and amplitude ($\epsilon_\eta/\Omega$) for the model
in Sec.~\ref{sin}.} \label{F(v,e)}
\end{figure}

We have also studied, with the same methods, the response of the
system in presence of analogous perturbations which have different
symmetries. We have considered the case in which only the real part
of (\ref{monoc_noise}) is taken; in this case the perturbation acts
only in the amplitude of the coupling but not in the de-tuning. We
have also analyzed the case of a perturbation which is square wave
shaped; in this case a {\it probe} function is identified by its
half period and initial phase. In both cases the corresponding
patterns of the average gate fidelity are exactly analogous to the
one shown in Fig.~\ref{F(v,e)}. This leads to the conclusion that
the pattern of fidelity is largely independent of the details of the
chosen probe function and a rather general behavior as function of
the typical frequency is observed.

Analogous results are also found for other loops of the same kind,
such as the loop with the angle $\varphi$ varying from $0$ to
$\pi/4$ in (\ref{path}) which is related to the Hadamard gate.

\subsection{\label{sphere}Random noise on the sphere}

In this section we consider a model of noise which preserves the
spectrum of the unperturbed Hamiltonian (\ref{Ham}), because of its
symmetries an analytical solution of the noisy dynamics is
available.

In~\cite{Florio_1} it was shown that the evolution operator can be
evaluated without approximation in several situations. Referring to
the model in Eq.~(\ref{Ham}), it is possible to evaluate the
evolution operator along \emph{any} segment on the parametric sphere
as far as one of the parameters $(\vartheta,\varphi)$ is kept
constant. In particular, referring to the case in
Fig.~\ref{loop}(b), the loop is composed by three segments and along
each of them the previous condition is satisfied. Thus one can
demonstrate~\cite{Florio_1} that the total evolution operator can be
splitted in the form
\begin{equation}
U(\tau)=U_{3}(\tau_3)\,U_{2}(\tau_2)\,U_{1}(\tau_1),
\end{equation}
where $\tau$ is the total time evolution and $\tau_i,\,i=1,2,3$ are
the times needed for covering each segment (for simplicity we
suppose that the speed of the evolution are constant in each
segment); moreover, the intermediate $U_i$'s can be explicitly
calculated~\cite{Florio_1}. Their form is very peculiar and it is
possible to see that, in terms of the parameters
$(\vartheta,\varphi)$, one can write
\begin{eqnarray}
U_{1}(t_1) & = & U_{1}(\vartheta_1(t_1),\varphi_1)\label{U1},\\
U_{2}(t_2) & = & U_{2}(\vartheta_2,\varphi_2(t_2))\label{U2},\\
U_{3}(t_3) & = & U_{3}(\vartheta_3(t_3),\varphi_3)\label{U3},
\end{eqnarray}
where $0\leq t_i \leq \tau_i$ and $\varphi_{1}$, $\vartheta_{2}$ and
$\varphi_{3}$ are the constant values of the parameters during the
evolution along the segment $1$, $2$ and $3$ respectively.

\begin{figure}
\centering
\includegraphics[width=0.48\textwidth]{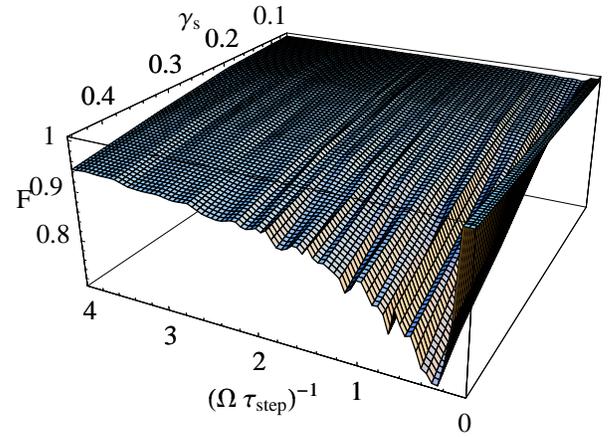}
\caption{(Color on line.) Average gate fidelity at the first
optimal operational time as a function of the re-scaled
adimensional noise amplitude $\gamma_s=\gamma/(\pi/2)$ and the
adimensional parameter ${(\Omega\tau_{\text{step}})}^{-1}$ for the
model in Sec.~\ref{sphere}.} \label{fidan}
\end{figure}

We want to use these results for gaining information about the
influence of the noise. We will therefore consider the following
model: every $U_i$ is splitted in $N$ evolution operators $U_i^j$
evolving for a time $\tau_{\text{step}}$ (a sub-segment) such that
\begin{equation}\label{numfluct}
N=\tau_i/\tau_{\text{step}}.
\end{equation}
The evolution in the segment $i$ reads
\begin{equation}\label{decompositionsegment}
U_i(\tau_i)=\prod_{j=1}^N U_i^j(\tau_{\text{step}}).
\end{equation}
In each sub-segment one of the sphere parameters is constant and
the other evolves (we are moving on meridians or parallels). We
add a random component to the constant parameter while the other
is not affected. In other words, we are including a transverse
component. We also suppose that the transverse evolution operator
is equal to the identity (the ``switch'' is infinitely fast). This
way we have splitted the evolution on a single meridian (parallel)
in a sequence of evolutions of shorter meridians (parallels).
Using Eq.s (\ref{U1})-(\ref{decompositionsegment}) we can write
\begin{eqnarray}
U_{1}(\tau_1)=\prod_{j=1}^N U_1^j(\vartheta_1(t_1^j),\varphi_1+\xi_1^j),\\
U_{2}(\tau_2)=\prod_{j=1}^N U_2^j(\vartheta_2+\xi_2^j,\varphi_2(t_2^j)),\\
U_{3}(\tau_3)=\prod_{j=1}^N
U_3^j(\vartheta_3(t_3^j),\varphi_3+\xi_3^j),
\end{eqnarray}
where $(j-1)\tau_{\text{step}} \leq t_i^j \leq j\tau_{\text{step}}$
and $\xi_i^j \in[-\gamma,\gamma]$ are random variables uniformly
distributed in the chosen interval ($i=1,2,3$). We stress again that
each operator in the decomposition has a (large and not transparent)
analytical expression. Using this model we have computed the average
gate fidelity at the first optimal operational time $\tau^*_1$ by
means of Eq.~(\ref{pauli_fid}) (and averaged over $50$ realizations
of the random process). The result is shown in Fig.~\ref{fidan}; $F$
is plotted as a function of the noise amplitude $\gamma$ (re-scaled
with the maximum value of the parameter for the loop in
Fig.~\ref{loop}(b) i.e.\ $\pi/2$) and the parameter
$(\Omega\tau_{\text{step}})^{-1}$ (characterizing the frequency of
the noise); notice that, due to Eq.~(\ref{numfluct}), at fixed value
of the operational time $\tau$, higher values of
$(\Omega\tau_{\text{step}})^{-1}$ correspond to a larger number of
fluctuations.
Also for this model the fidelity exhibits a breakdown for small
frequencies of the noise (which is in accordance with previous
results). In particular, for
${(\Omega\tau_{\text{step}})}^{-1}<0.5$ the fidelity exhibits a
minimum for any amplitude of the noise. Anyway, we notice that the
deep of the fidelity is pronounced if the noise amplitude is one
half the maximum value of the parameters; clearly, this situation
corresponds to an unphysical scenario in which the control of the
parameters is very poor. In all other situations the typical
values of $F$ is very high. In the range of intermediate and large
frequencies $F$ quickly recovers the ideal behavior.

\begin{figure}
\centering
\includegraphics[width=0.48\textwidth]{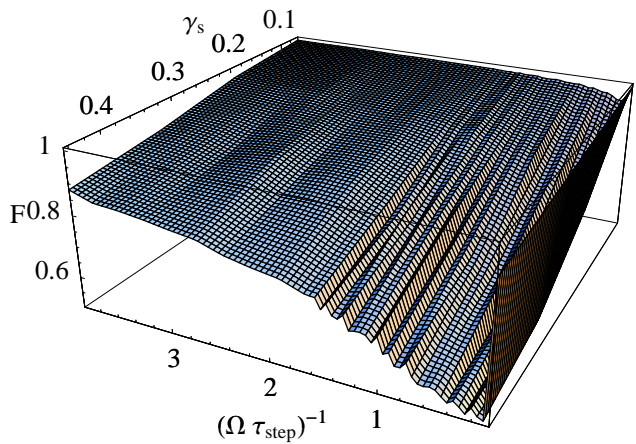}
\caption{(Color on line.) Average gate fidelity at the fourth
optimal operational time as a function of the re-scaled noise
amplitude $\gamma_s=\gamma/(\pi/2)$ and the adimensional parameter
${(\Omega\tau_{\text{step}})}^{-1}$ for the model in
Sec.~\ref{sphere}.} \label{faa}
\end{figure}

It is interesting to compare the behavior of the gate at the first
optimal operational time to the case of longer operational time in
presence of noise, i.e., in the (approximated) adiabatic regime. It
is possible to see~\cite{Florio_1,Florio_2} that the fidelity
oscillations shown in Fig.~\ref{F(T)} in absence of noise are
strongly suppressed if $k\ge 3$ in Eq.~(\ref{opttimes}) (we are near
the adiabatic regime). A good approximation of the adiabatic regime
can be already obtained for the fourth optimal operational time.
Therefore, we have computed the average gate fidelity for
$\Omega\tau^*_4 \simeq 75.21$ [${(\Omega\tau_{\text{step}})}^{-1}$
ranges as in Fig.~\ref{fidan}]. The result is shown in
Fig.~\ref{faa} and can be directly compared to the plot in
Fig.~\ref{fidan}. First of all it is important to stress that the
total number of fluctuations $N_4$ for $\Omega\tau^*_4 \simeq 75.21$
is larger when compared to the number $N_1$ for the first optimal
working point $\Omega\tau^*_1 \simeq 18.25$. From Eq.
(\ref{opttimes}) and Eq.~(\ref{numfluct}) we have $N_4\simeq 4.12
N_1$. In apparent contrast to the intuition related to the usual
argument of robustness of holonomic gates we notice that, in the
same range of frequencies of the non adiabatic case (and, therefore,
for a larger number of fluctuations), $F$ reaches lower values.
Moreover, the adiabatic gate needs higher values of the frequency of
noise for recovering the ideal behavior. We conclude that the
(approximately) adiabatic (purely geometric) NOT transformation is
more sensitive to parametric noise than the non adiabatic one.

\subsection{\label{random}Random noise}

In this section we consider a model for a random perturbation of the
loop which is not constrained to preserve the sphere in the
parameter space. Taking in consideration the ideal loop (\ref{path})
here we study the noisy paths of the following kind:
\begin{eqnarray}
\left\{
\begin{array}{ccc}
x_\mathrm{n}(s;\tau_\text{step},\epsilon,\tau) & = & x(s) + \xi_1(s,\tau_\text{step},\tau) \\
y_\mathrm{n}(s;\tau_\text{step},\epsilon,\tau) & = & y(s) + \xi_2(s,\tau_\text{step},\tau) \\
z_\mathrm{n}(s;\tau_\text{step},\epsilon,\tau) & = & z(s) +
\xi_3(s,\tau_\text{step},\tau)
\end{array} \right.\;,
\end{eqnarray}
where $\xi_i(s,\tau_\text{step},\tau) \in [-\epsilon,\epsilon]$ are
three real random variables, uniformly distributed in the chosen
interval, which are piecewise constant for $(j-1)\tau_\text{step}
\leq s\tau \leq j\tau_\text{step}$.

In order to study the behavior of the gate at finite operational
time, we have evaluated the average gate fidelity for a fixed value
of the noise amplitude $\epsilon=0.1\Omega$ as a function of the
noise typical frequency $(\Omega\tau_\mathrm{step})^{-1}$ in
correspondence of the first four optimal working times.
\begin{figure}
\centering
\includegraphics[width=0.48\textwidth]{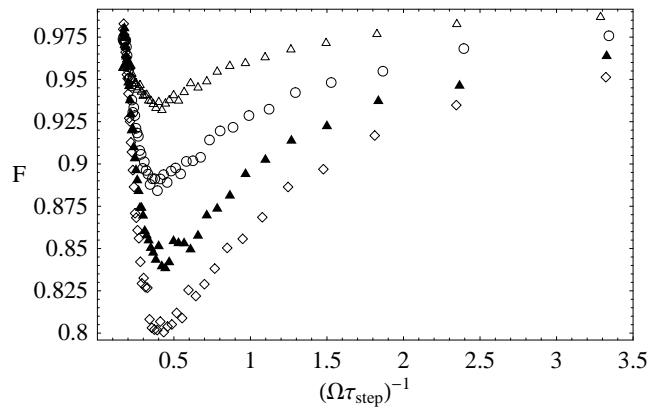}
\caption{Average gate fidelity as a function of the noise typical
frequency for the noise model in Sec.~\ref{random} for the first
four optimal operational times. Triangles, circles, full triangles
and squares correspond respectively to the first, second, third and
fourth optimal operational time. $\epsilon = 0.1 \Omega$.}
\label{collapse}
\end{figure}
The results are shown in Fig.~\ref{collapse}. The data plotted in
this figure lead us to two considerations: first of all we notice
again the unexpected result that the non-adiabatic optimal working
times (the first, for instance) appears to be more robust than
longer operational times (the forth optimal operational time, for
instance); secondly, we observe the same qualitative behavior of the
pattern of fidelity for all the optimal operational times under
study, this suggests the presence of a common mechanism which
account for the cancellation of the effects of the noise.

We have also analyzed the case of a noise which include de-tuning
by considering complex random variables
$\xi_i(s,\tau_\text{step},\tau)$. The result are completely
analogous and the introduction of a noise in the de-tuning does
not introduce new elements in the pattern of fidelity.

\section{\label{kritik}Analysis of the results}

The aim of this section is to look for a physical explanation of the
observed robustness of the considered finite time non-adiabatic
gate. Due to the fact that all the models of noise induce the same
qualitative behavior of the fidelity, in the following we are going
to consider in more details the model presented in
Sec.~\ref{random}. It is worth to notice that only for the first
model the noise affects both the amplitude and the phase of the
control parameters, while the other two models of noise concern only
their amplitude. Nevertheless, it is a result
of~\cite{Solinas_Zanardi} that the main contribution in the noisy
dynamics is due to the component in the amplitude.

\begin{figure}
\centering
\includegraphics[width=0.48\textwidth]{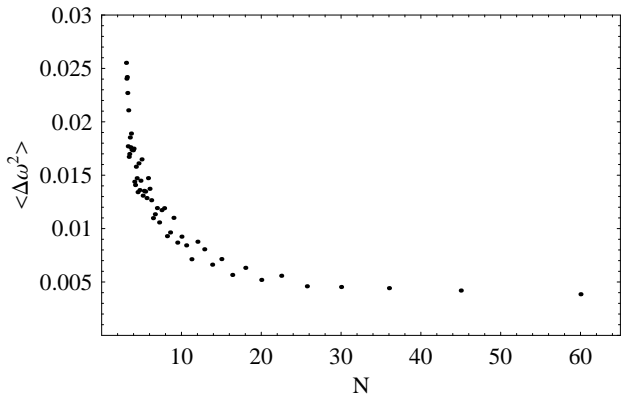}
\caption{Fluctuations in the solid angle spanned by a noisy loop
as a function of the number of perturbations of the noise $N$, for
the noise model in Sec.~\ref{random}. $\epsilon=0.1\Omega$.}
\label{angle}
\end{figure}

As already recalled, the relevant parameter for the geometrical
cancellation usually related to holonomic gates in the adiabatic
regime is the number of fluctuations of the noise during the gate
operational time (denoted $N$). This effect is related only to the
swept solid angle and is independent of the chosen operational time.
If the number of cycles of the noise is large enough, the
fluctuations in the solid angle spanned by the loop are expected to
become negligible. To be more specific suppose that, after a noisy
loop, the swept solid angle is $\omega$ and the mean square over the
realizations of the noise is $\langle \Delta\omega^2 \rangle$. In
Fig.~\ref{angle} the mean square is plotted as a function of the
number of cycles of the noise; since, in the adiabatic limit, the
gate depends only on the swept solid angle, the fluctuations of the
gate are expected to have the same behavior as the fluctuations in
the solid angle.

\begin{figure}
\centering
\includegraphics[width=0.48\textwidth]{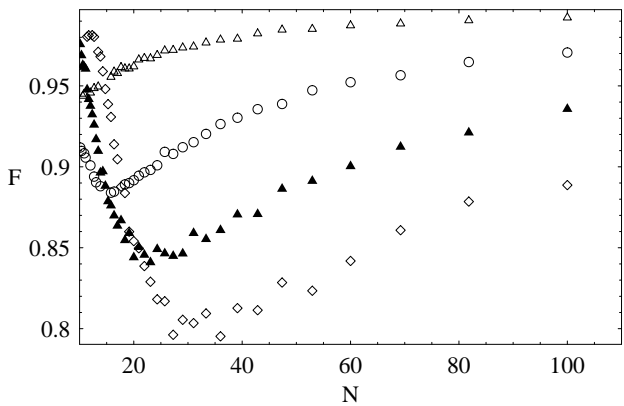}
\caption{Average gate fidelity as a function of the number of
fluctuations of the noise $N$ for the noise model in
Sec.~\ref{random} for the first four optimal operational times.
Triangles, circles, full triangles and squares correspond
respectively to the first, second, third and fourth optimal
operational time. $\epsilon=0.1\Omega$. Compare with
Fig.~\ref{collapse} and \ref{angle}.} \label{no-collapse}
\end{figure}

As already explained in the previous section, Fig.~\ref{collapse}
shows the average gate fidelity as a function of the adimensional
typical noise frequency $(\Omega\tau_{\mathrm{step}})^{-1}$ for
several values of the evolution time which correspond to the first
four optimal operational times. The plot shows an analogous behavior
of the fidelity as a function of the typical noise frequency
\emph{independently} of the particular value of the operational
time; moreover, the minimum of the fidelity is reached in
correspondence of $(\Omega\tau_{\mathrm{step}})^{-1}\simeq 0.5$ for
all the values of the operational time considered. In order to cast
some light on the nature of the cancellation effect, the same data
are plotted in Fig.~\ref{no-collapse} as functions of the number of
fluctuations of the noise (notice that
$N=\Omega\tau(\Omega\tau_{\mathrm{step}})^{-1}$).
A direct comparison of figures \ref{collapse} and \ref{no-collapse}
suggests that the relevant quantity which accounts for the mechanism
of cancellation of the effects of the noise is its typical frequency
$(\Omega\tau_{\mathrm{step}})^{-1}$ and \emph{not only} the number
of fluctuations $N$. On the other hand, the fluctuations of the
solid angle around the ideal value ($\pi/2$) start to be negligible
for $N>20$; a comparison with the curve for the fourth optimal
working point (squares in Fig.~\ref{no-collapse}) suggests that the
recovery of the fidelity for long cyclic evolution times is given
also by geometric cancellation.

For non adiabatic evolution times one can imagine the existence of a
different mechanism which accounts for the observed cancellation of
the noise effects for sufficiently fast noise which is related to a
\emph{dynamical} instead of geometrical cancellation. A dynamical
effect could not be directly related to the swept solid angle: in
this case the relevant parameter is expected to be the typical time
of the noise $\tau_{\mathrm{step}}$ and a dynamical cancellation of
the noise should appear if its typical frequency is sufficiently
large compared to the system frequency, namely
$(\Omega\tau_{\mathrm{step}})^{-1}\gg 1$.
Of course this condition implies, for fixed operational time $\tau$,
that $N \gg 1$ (the usual condition for geometric cancellation);
nevertheless, as Fig.~\ref{collapse} shows, a cancellation of the
noise effects appears on a frequency scale
$(\Omega\tau_{\mathrm{step}})^{-1} \simeq 1$ \emph{independently} of
the chosen value of the operational time, thus suggesting a
dynamical mechanism for the noise cancellation at least for the
first four optimal operational times.

The fact that in the non adiabatic regime the robustness has a
dynamical origin can also explain why the minimum value of the
fidelity tends to decrease for increasing values of $\tau^*$: if the
geometric cancellation is not present, the noise is less effective
in disturbing the system when the evolution time is short.

\section{\label{conclude}Conclusions}

In this paper we have considered the influence of parametric noise
on the efficiency of a non adiabatic holonomic gate which is known
to be robust in the ideal case. Three models of parametric noise
have been discussed in the case of finite operational time. The
average gate fidelities for all the models of noise considered
here present an analogous qualitative behavior. For each of the
three models the non ideal gate presents a breakdown of the
average gate fidelity for small frequencies of the noise (compared
to the system Bohr frequency), while a high value of the fidelity
is reached for noise with higher frequencies. This can lead to say
that the presence of a ``resonant frequency'' for the breakdown of
$F$ is a general feature of any model of parametric noise.

We want to stress again that the usual argument in favor of the
robustness of holonomic quantum computation is based on the purely
geometric nature of the holonomy group that describes the
adiabatic transformations. Since the dynamics has a
\emph{completely} geometric character \emph{only} in the adiabatic
limit, the robustness of adiabatic gates is, in this sense, just a
consequence of the adiabatic theorem. Despite these
considerations, our calculations show that, at least in certain
situations, the first optimal operational time can be preferable
to longer operational times with regards to the robustness of the
corresponding gate against parametric noise.

Nevertheless, our results lead to the conclusion that the observed
revivals of the fidelity for sufficiently fast noises is mainly due
to \emph{dynamical} instead of geometrical effects. Our conclusion
is that, in the range of operational times considered here, the
observed cancellation effects are mainly related to a dynamical
average over fast oscillations of the noise
$(\Omega\tau_{\mathrm{step}})^{-1} \gg 1$ and there is no relevant
connection with the swept solid angle which plays a crucial role for
the usual argument in favor of robustness of the holonomic
computation in the adiabatic regime.

\acknowledgments This work is partly supported by the European
Community through the Integrated Project EuroSQIP and by the
bilateral Italian--Japanese Projects II04C1AF4E on ``Quantum
Information, Computation and Communication'' of the Italian
Ministry of Education, University and Research. G.F. thanks Paolo
Facchi, Saverio Pascazio and Gianni Costantini for useful
discussions and the Quantum Transport Group at TU Delft for kind
hospitality and support.


\begin{thebibliography}{99}

\bibitem{Nielsen_Chuang} M. A. Nielsen, I. L. Chuang,
\textit{Quantum Computation and Quantum Information} (Cambridge
University Press, Cambridge, 2000).

\bibitem{Casati} G. Benenti, G. Casati and G. Strini,
\textit{Principles of Quantum Computation and Information} (World
Scientific, Singapore, 2004).

\bibitem{Breuer} H. P. Breuer, F. Petruccione,
\textit{The Theory of Open quantum Systems} (Oxford University
Press, Oxford, 2002).

\bibitem{nonMarkov} R. Alicki, M. Horodecki, P. Horodecki, R. Horodecki, L. Jacak and P.
Machnikowski, {\bf 70}, 010501(R) (2004);
K. Roszak, A. Grodecka, P. Machnikowski and T. Kuhn,
Phys. Rev. B {\bf 71}, 195333 (2005).

\bibitem{NIST} D. J. Wineland, C. Monroe, W. M. Itano, D. Leibfried,
B. E. King, D. M. Meekhof, J. Res. Natl. Inst. Stand. Technol.
{\bf 103}, 259 (1998).

\bibitem{Zanardi_Rasetti} P. Zanardi, M. Rasetti,
Phys. Lett. A {\bf 264}, 94 (1999).

\bibitem{PachosPRA61} J. Pachos, P. Zanardi and M. Rasetti,
Phys. Rev. A {\bf 61}, 010305(R) (1999).

\bibitem{Nakahara} M. Nakahara,
\textit{Geometry, topology and physics}, 2nd Ed., (IoP publishing,
Bristol, 2005).

\bibitem{Marathe} K. B. Marathe, G. Martucci,
\textit{The Mathematical Foundations of Gauge Theories}
(North-Holland, Amsterdam, 1992).

\bibitem{Berry} M. V. Berry,
Proc. R. Soc. Lond. A {\bf 392}, 45 (1984); B. Simon,
Phys. Rev. Lett. {\bf 51}, 2167 (1983).

\bibitem{WZ} F. Wilczek and A. Zee,
Phys. Rev. Lett. {\bf 52}, 2111 (1984).

\bibitem{Pachos} J. Pachos, P. Zanardi,
Int. J. Mod. Phys. B {\bf 15}, 1257 (2001).

\bibitem{Kato} T. Kato,
J. Phys. Soc. Jap. {\bf 5}, 435 (1951).

\bibitem{Aharonov} Y. Aharonov, J. Anandan,
Phys. Rev. Lett. {\bf 58},  1593 (1987).

\bibitem{Anandan} J. Anandan,
Phys. Lett. A {\bf 133}, 171 (1988).

\bibitem{DeChiara_Palma} G. De Chiara, G. M. Palma,
Phys. Rev. Lett. {\bf 91}, 090404 (2003).

\bibitem{Solinas_Zanardi} P. Solinas, P. Zanardi and N. Zangh\`i,
Phys. Rev. A {\bf 70}, 042316 (2004).

\bibitem{Zu_Zanardi} S.-L. Zhu and P. Zanardi,
Phys. Rev. A {\bf 72}, 020301(R) (2005).

\bibitem{Carollo} A. Carollo, I. Fuentes-Guridi, M. Fran\c{c}a Santos
and V. Vedral,
Phys. Rev. Lett. {\bf 90}, 160402 (2003).

\bibitem{Carollo2} A. Carollo, I. Fuentes-Guridi, M. Fran\c{c}a Santos
and V. Vedral,
Phys. Rev. Lett. {\bf 92}, 020402 (2004).

\bibitem{Fuentes} I. Fuentes-Guridi, F. Girelli, E. Livine,
Phys. Rev. Lett. {\bf 94}, 020503 (2005).

\bibitem{Parodi} D. Parodi, M. Sassetti, P. Solinas, P. Zanardi, N.
Zangh\`i,
Phys. Rev. A {\bf 73}, 052304 (2006).

\bibitem{Wu_Zanardi_Lidar} L.-A. Wu, P. Zanardi, D. A. Lidar,
Phys. Rev. Lett. {\bf 95}, 130501 (2005).

\bibitem{Florio_1} G. Florio, P. Facchi, R. Fazio, V. Giovannetti and S.
Pascazio,
Phys. Rev. A {\bf 73}, 022327 (2006).

\bibitem{Florio_2} A. Trullo, P. Facchi, R. Fazio, G. Florio, V. Giovannetti, S.
Pascazio,
Las. Phys. {\bf 16}, 1478 (2006).

\bibitem{Florio_3} G. Florio,
Open Sys. \& Information Dyn. {\bf 13}, 263 (2006).

\bibitem{Cirac} L.-M. Duan, J. I. Cirac, P. Zoller,
Science {\bf 292}, 1695 (2001).

\bibitem{Falci} G. Falci, R. Fazio, G. M. Palma, J. Siewert and V.
Vedral, Nature {\bf 407}, 355 (2000).

\bibitem{Faoro} L. Faoro, J. Siewert and R. Fazio, Phys. Rev. Lett.
{\bf 90}, 028301 (2003).

\bibitem{Solinas} P. Solinas, P. Zanardi, N. Zangh\`i and F. Rossi,
Phys. Rev. A  {\bf 67}, 062315 (2003).

\bibitem{SL} M. S. Sarandy, D. A. Lidar,
Phys. Rev. Lett. {\bf 95}, 250503 (2005).

\bibitem{Sarandy_Lidar} M. S. Sarandy, D. A. Lidar,
Phys. Rev. A {\bf 71}, 012331 (2005);
{\bf 73}, 062101 (2006).

\bibitem{Born_Fock} M. Born and V. Fock,
Z. Phys. {\bf 51}, 165 (1928).

\bibitem{Nielsen} M. A. Nielsen,
Phys. Lett. A {\bf 303}, 249 (2002); M. D. Bowdrey, D. K. L. Oi,
A. J. Short, K. Banaszek and J. A. Jones, {\it ibid.} {\bf 294},
258 (2002); M. Horodecki, P. Horodecki and R. Horodecki, Phys.
Rev. A {\bf 60}, 1888 (1999).

\end{thebibliography}
\end{document}